\documentclass[showpacs,prd,aps]{revtex4}
\usepackage{amsfonts}
\usepackage{amsmath}
\usepackage[dvips]{graphicx}
\usepackage{epsfig}

\begin{document}

\title{Pre-Holography}

\author{Bernard~S.~Kay\footnote{Electronic address: bsk2@york.ac.uk} and
Peter~Larkin\footnote{Electronic address: pcl500@york.ac.uk}}

\affiliation{Department of Mathematics, University of York, York YO10 5DD, U.K.}

\begin{abstract}

\noindent
We construct a symplectic isomorphism, $\mathfrak{h}$, from classical Klein
Gordon solutions of mass $m$ on $(d+1)$-dimensional Lorentzian Anti de
Sitter space (equipped with the usual symplectic form) to a certain
symplectic space of functions on its conformal boundary (only) for all
integer and half-integer $\Delta$
($=\frac{d}{2}+\frac{1}{2}(d^{2}+4m^{2})^{1/2}$). 
$\mathfrak{h}$ induces a large family of new examples of Rehren's
\textit{algebraic holography} in which the net of local quantum Klein
Gordon algebras in AdS is seen to map to a suitably defined net of local
algebras for the (generalized free) scalar conformal field with
anomalous dimension $\Delta$ on $d$-dimensional Minkowski space (the AdS
boundary).  Relatedly, we show for these models that Bertola et al's
\textit{boundary-limit holography} becomes a quantum duality (only) if
the test functions for boundary Wightman distributions are restricted
in a particular way.

\end{abstract}

\pacs{04.62.+v, 11.10.Cd, 11.25.Tq, 11.25.Hf}

\maketitle
 
The conjecture \cite{Maldacena:1997re} in 1997 of a holography-like
correspondence between a certain type of string theory on the bulk of
Anti de Sitter space (AdS) (in 5-dimensions and producted with a 5-sphere)
and a certain limit of a certain family of
conformal field theories (CFT) on its (conformal) boundary (or between
supergravity on the bulk and the same limit of CFT on the boundary
\cite{Gubser:1998bc, Witten:1998qj}) has led to many new and surprising
conjectured interrelationships between quantum gravity and Minkowskian
quantum field theory.  One spin-off of this conjecture was that a number
of authors (see especially  \cite{RehAH, Bertola:2000pp}) began to
investigate the related, but distinct and simpler, question: In what
sense can a correspondence be established between an ``ordinary'' (e.g.
scalar) quantum field theory on a (Lorentzian, $(d+1)$-dimensional) AdS background and a
suitable ``ordinary'' (conformal) field theory on its conformal
boundary?  This is a simpler question because it concerns not full
quantum gravity but quantum field theory in curved spacetime \cite{Kay:2006jn}. 
Two different sorts of answer to this question were proposed, the
\textit{algebraic holography} of Rehren \cite{RehAH} and  the
\textit{boundary-limit holography} of Bertola-Bros-Moschella-Schaeffer
\cite{Bertola:2000pp} -- both in the context of axiomatic quantum field
theory \cite{Haag}. 

Rehren's algebraic holography \cite{RehAH} is formulated in terms of the
algebraic version of axiomatic quantum field theory.  In this framework,
the specification of a given quantum field theory on a given background
spacetime is tantamount \cite{Kay:2006jn} to the specification of a
\textit{net of local $*$-algebras}.  In other words, the specification,
for each (suitable) region $\cal O$ of the background spacetime, of a
$*$-algebra $\cal A(\cal O)$ -- the collection of the latter algebras
being \textit{isotonous} which means that when one region sits inside
another, then its algebra is a subalgebra of the algebra of the larger
region. 

The basic idea of algebraic holography is to map a given spacelike
wedge (defined as in \cite{RehAH}) in AdS to its intersection with the
boundary.  As Rehren points out, this sets up a bijection between the set
of all wedges in the bulk and the set of all double-cones on the
boundary which moreover maps spacelike related bulk wedges to spacelike
related boundary double-cones \cite{ftnt1}.  If we are then given a net
of local algebras on the bulk (where, in our definition above,
``region'' is interpreted to mean wedge) then  algebraic holography
consists of the definition of a net of local algebras on the boundary
(where, in our definition above, ``region'' is interpreted to mean
double-cone) by identifying the algebra for a given boundary double-cone
with the bulk wedge algebra which restricts to it \cite{ftnt1, ftnt2}.
    
Bertola et al's boundary-limit holography is formulated in terms of the
Wightman version of axiomatic quantum field theory. In this framework,
and assuming the theory involves only a single scalar field, the
specification of a given quantum field theory on a given background
spacetime is tantamount to the specification of a family of Wightman 
distributions  $W_n(f_1, \dots, f_n)$ for each integer $n$, each
of which may roughly be  interpreted as the result of smearing the
(singular)  $n$-point  ``expectation value'' $\langle 0|\phi(x_1), \dots
\phi(x_n)|0\rangle$ in a suitable ``vacuum state'' $|0\rangle$, with
(smooth, compactly supported)  test functions $f_1\dots f_n$.  In an
oversimplified description, where one ignores the need to smear, what
Bertola et al show may be described by saying that, for a given family
of Wightman functions $W((t_1, \rho_1, \Omega_1), \dots ,(t_n,
\rho_n, \Omega_n))$ in the bulk of AdS,  if one chooses $\Delta$
suitably, then the limit $\lim_{\rho_1,\dots ,\rho_n\rightarrow
\pi/2}(\cos\rho_1\dots\cos\rho_n)^{-\Delta} W((t_1, \rho_1,
\Omega_1), \dots ,(t_n, \rho_n, \Omega_n))$ will exist and define a
family of Wightman functions  $W((t_1, \Omega_1), \dots ,(t_n,
\Omega_n))$ on the conformal boundary which belong to a CFT.  What they
actually show is that a correct distributional counterpart to this
limiting procedure maps any Wightman theory in the bulk to a Wightman
theory for the appropriate CFT on the boundary.

Above, we have used the usual global coordinates in which the
AdS$_{d+1}$ metric takes the form\hfil\break 
$ds^{2}=\sec^{2}\rho dt^{2}-\sec^{2}\rho d\rho^{2}-\tan^{2}\rho
d\Omega_{d-1}^{2}$ with $0\leq\rho\leq\pi/2$, $-\infty\leq t\leq\infty$ and
$\Omega$ denotes the usual angular coordinates on the $(d-1)$-sphere. 
In the special case of the Klein Gordon equation (KG), 
\begin{equation}
\label{KG}
(\cos^{2}\rho\partial^{2}_{t}-\cos^{2}\rho\partial^{2}_{\rho}-
(d-1)\cot\rho\partial_{\rho}-\cot^{2}\rho\nabla_{S^{d-1}}^{2}+m^{2})
\phi(t,\rho,\Omega)=0 \end{equation} quantized on AdS according to the  
Avis-Isham-Storey scheme \cite{Avis:1977yn} for vanishing boundary
conditions, one finds that the two-point distribution in the bulk  has a
non-trivial boundary-limit when $\Delta$ takes the value (cf.
\cite{Witten:1998qj}) \begin{equation} \label{deltadef}
\Delta=\frac{d}{2}+\frac{1}{2}(d^{2}+4m^{2})^{1/2} \end{equation} which,
when one identifies the appropriate part of the boundary (see endnote
\cite{ftnt1}) with $d$-dimensional Minkowski space, turns out to
transform to the standard two-point function  \begin{equation}
\label{Wb} W_b(x,x')=\frac{1}{2\pi^{\frac{d}{2}}}\frac{\Gamma(\Delta)} 
{\Gamma(\Delta-d/2+1)}\frac{1} {[-(t-t'-i\epsilon)^{2}+({\bf x}-{\bf
x'})^2]^\Delta} \end{equation} for a conformal scalar field,
$\hat\phi_d^\Delta$, of anomalous dimension $\Delta$ (and other
$n$-point functions will be those of a generalized free field with this
2-point function).

The work we report here had two interrelated purposes: to use the bulk
KG model to construct examples of algebraic holography and to clarify
the relation between boundary-limit and algebraic holography.  As we
shall see below, whenever $\Delta$ (\ref{deltadef}) is an integer or
half-integer, we have found a way to fulfill both of these purposes and
we will show, first, that, for such $\Delta$, if one starts with the net
of local algebras for a bulk KG field, then the net of local algebras
defined on the boundary by algebraic holography coincides with the
subnet of local algebras for $\hat\phi_d^\Delta$ which results when one 
replaces the usual test functions by a certain smaller family of test
functions and ``localizes'' them in a suitable way as we will explain
and discuss below.  Second, we show that, for the same $\Delta$, if one
restricts the range of the Bertola et al projection to the Wightman
functions of $\hat\phi_d^\Delta$ smeared only with the same smaller
family of test functions, then the resulting quantum theory is dual
(i.e. isomorphic) to the bulk quantum theory.  

\begin{figure}[htbp]
\centering		
\includegraphics[width=0.30\textwidth]{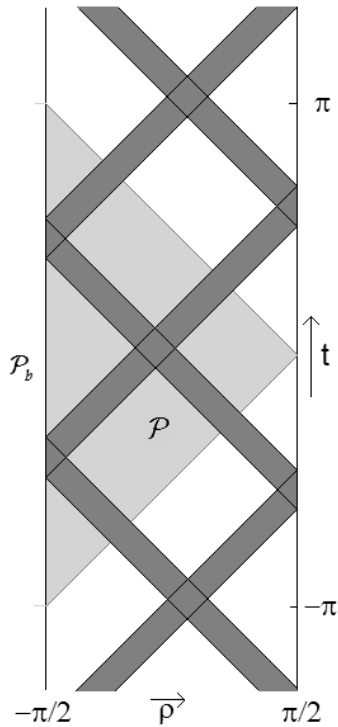}
\label{fig: zig_zag}
\caption{The support (dark shading) of a ``typical'' classical solution
for the bulk massless scalar field on AdS$_{1+1}$ and a choice (light
shading) of chart $\cal P$ for Poincar\'e coordinates. (See endnote
\cite{ftnt3}.)}
\end{figure}

In order to obtain these results, we import into, and adapt to this
AdS-CFT context, the mathematical formalism (see \cite{Kay:2006jn})
which has been successful in constructing and analysing the properties
of linear quantum fields in other curved spacetime contexts.  The key to
everything we do is the construction, for the KG equation on AdS,
whenever $\Delta$ is an integer or half-integer, of a classical
counterpart to quantum holography, which we call the 
\textit{pre-holography} map $\mathfrak{h}$.  

To construct this, we first introduce the space, $S$, of smooth
classical solutions to (\ref{KG}) on AdS$_{d+1}$ which vanish on the
conformal boundary. We recall that (for $d\ge 2$ \cite{ftnt3}) any such
classical solution may be expanded \cite{Balasubramanian:1999ri} as
\begin{equation}
\label{phibulkexpansion}
\phi(t,\rho,\Omega)=
\sum_{nl\vec{m}}\left(\frac{\Gamma(1+n)\Gamma(\Delta+l+n)}
{\Gamma(l+\frac{d}{2}+n)\Gamma(1+\Delta-\frac{d}{2}+n)}\right)^\frac{1}{2}\sin^{l}
\rho\cos^{\Delta}\rho
P_{n}^{(l+\frac{d}{2}-1,\Delta-\frac{d}{2})}(\cos 2\rho)
(a_{nl\vec{m}}e^{-i(\Delta+l+2n)t}Y_{l\vec{m}}(\Omega) + {\rm c.c})  
\end{equation}
where $P_n^{(\alpha,\beta)}(x)$ are Jacobi polynomials \cite{Bateman},
$Y_{l,\vec{m}}$ are the ($L^2$-normalised) spherical harmonics on the
$(d-1)$-sphere and the sum is over $n$ from 0 to $\infty$ and the usual
ranges of $l$ and $\vec{m}$. We equip $S$ with the (standard
\cite{Kay:2006jn}) symplectic form 
$\sigma(\phi_{1},\phi_{2})=\int_{t={\rm const}}
(\phi_{1}\dot{\phi}_{2}-\dot{\phi}_{1}\phi_{2})g^{00}\sqrt{g}d^{d}x$
$=\sum_{nl\vec{m}}i(a_{nl\vec{m}}^{1}a_{nl\vec{m}}^{2*}-a_{nl\vec{m}}^{1*}
a_{nl\vec{m}}^{2})$ where the integral is over any $t={\rm const}$ surface.
We then define our pre-holography map, $\mathfrak{h}$, to be the map which sends
such a classical solution to the function on the conformal boundary
which has the expansion
\begin{equation}
\label{phibsym}
\phi_b^s(t,\Omega) = 
\sum_{nl\vec{m}}\frac{i\Gamma(\Delta-\frac{d}{2}+1)}{\pi}\left 
({\frac{\Gamma(l+\frac{d}{2}+n)n!}{\Gamma(\Delta+l+n)
\Gamma(n+\Delta-\frac{d}{2}+1)}}\right )^\frac{1}{2}(a_{nl\vec{m}}
e^{-i(\Delta+l+2n)t}Y_{l\vec{m}}(\Omega) - {\rm c.c})
\end{equation}
and we equip the range of $\mathfrak{h}$, which we call ${\cal F}_b$, with the 
antisymmetric bilinear form
\begin{equation}\label{sym_b}
\sigma_{b}({\phi_{b}^{s}}_{1},{\phi_{b}^{s}}_{2})=
\int\int dt_{1}dt_{2}\int\int d\Omega_{1}d\Omega_{2} 
E_{b}(t_{1},\Omega_{1};t_{2},\Omega_{2}) 
{\phi_{b}^{s}}_{1}(t_{1},\Omega_{1}){\phi_{b}^{s}}_{2}(t_{2},\Omega_{2})
\end{equation}
where the integration is over a choice \cite{ftnt1} (it obviously
doesn't matter which) of ${\cal P}_b$ region and $E_b$ is the
boundary-limit of the bulk Lichn\'erowicz (advanced minus retarded)
fundamental solution $E$:
\begin{eqnarray}
\label{eebee}
E_b(t_{1},\Omega_{1};t_{2},\Omega_{2})&=&\substack{\lim\\\rho_{1},\rho_{2}
\rightarrow\pi/2} 
(\cos\rho_{1}\cos\rho_{2})^{-\Delta}E(t_{1},\rho_{1},\Omega_{1};t_{2},\rho_{2},
\Omega_{2})\\\nonumber
&=&2{\rm Im}\sum_{nl\vec{m}}\frac{\Gamma(\Delta+l+n)\Gamma(n+\Delta-\frac{d}{2}+1)}
{\Gamma(l+\frac{d}{2}+n)n!\Gamma(\Delta-\frac{d}{2}+1)^{2}}
e^{-i(\Delta+l+2n)(t_{1}-t_{2})}Y_{l\vec{m}}(\Omega_{1})
Y_{l\vec{m}}^*(\Omega_{2})\nonumber.
\end{eqnarray}
We note in passing that 
$iE_b(x,x')I=2i{\rm Im}({\cal W}_b(x,x'))$ $=$ (when restricted to a ${\cal
P}_b$ region \cite{ftnt1}) $[\hat\phi_d^\Delta(x), \hat\phi_d^\Delta(x')]$. 

One can check, for any pair of classical solutions, $\phi_1$, $\phi_2$,
in $S$ with coefficients in their mode expansions denoted
$a_{nl\vec{m}}^{1}$, $a_{nl\vec{m}}^2$, that, defining
${\phi_b^s}_1$, ${\phi_b^s}_2$ as in equation (\ref{phibsym}), we have,
when (and only when) $\Delta$ is an integer or half-integer,
\begin{equation}
\label{symplecticpi}
\sigma_b({\phi_b^s}_1, {\phi_b^s}_2)
=\sum_{nl\vec{m}}i(a_{nl\vec{m}}^{1}a_{nl\vec{m}}^{2*}-
a_{nl\vec{m}}^{1*}a_{nl\vec{m}}^{2})=\sigma({\phi}_1, {\phi}_2)
\end{equation}
and thus, by the equality of the first and last expressions here, we
conclude both that $\sigma_b$ is non-degenerate, and hence a symplectic
form, and $\mathfrak{h}: S\rightarrow {\cal F}_b$ is a symplectic
isomorphism. The origin of the restriction to integer or half-integer
$\Delta$  lies in the calculation which is needed to show the first
equality in (\ref{symplecticpi}):  As may easily be seen, this
calculation involves integrals of form $\int_{-\pi}^\pi
\exp(\pm i(N+2\Delta)t)dt$ where $N$ is a positive integer and, for the
equality to hold, these integrals have to vanish and therefore $2\Delta$
has to be an integer.

We remark that the formula (\ref{sym_b}) may be written
\begin{equation}
\label{sym_bslick}
\sigma_b({\phi_b^s}_1,{\phi_b^s}_2)=\langle{\phi_b^s}_1|E_b*{\phi_b^s}_2\rangle
\end{equation}
where $\langle\cdot|\cdot\rangle$ denotes the $L^2$ inner product on our
choice of ${\cal P}_b$ region on the conformal boundary and $*$ denotes
convolution (i.e. smearing $E_b$ in its second argument).

A propos of $\sigma_b$ being non-degenerate, we remark that, when
restricted to a choice of ${\cal P}_b$, our space ${\cal F}_b$ falls
short \cite{ftnt3} of being the set of all smooth functions on ${\cal P}_b$ due to
the incompleteness of the set of modes in terms of which  $\phi_b^s$ is
expanded in (\ref{phibsym}).  Concomitantly, if we were to extend
$\sigma_b$ (restricted to ${\cal P}_b$) from the range, ${\cal
F}_b|_{{\cal P}_b}$ to the full set of smooth functions on ${\cal P}_b$,
then it would be degenerate since, due to the incompleteness of the set of modes in
terms of which it is expanded in (\ref{eebee}), the operator $E_b*$ has a
non-trivial kernel.    

Our purpose next is to exploit our just-defined pre-holography map,
$\mathfrak{h}$, to construct a mathematical object which corresponds to the
quantum boundary limit, $\hat\phi_b$, of the quantum bulk field,
$\hat\phi$, defined by the formal relation
\begin{equation}
\label{blimq}
\hat\phi_b(t,\Omega)=\lim_{\rho\rightarrow \pi/2} (\cos\rho)^{-\Delta}
\hat\phi(t,\rho,\Omega).  
\end{equation}
We know \cite{Kay:2006jn} the quantum bulk field, $\hat\phi$, can be
defined in terms of quantities  ``$\sigma(\hat\phi, \psi)$'' which
deserve to be considered the ``quantum bulk field, $\hat\phi$,
\textit{symplectically smeared} with a classical test solution $\psi$''
and which satisfy the commutation relations
\begin{equation}
\label{CCR}
[\sigma(\hat\phi, \psi_1), \sigma(\hat\phi, \psi_2)]=i 
\sigma(\psi_1, \psi_2)I, 
\end{equation}
and what we will do is to define, in terms of these $\sigma(\hat\phi,
\psi)$, a quantity which deserves to be called
``$\langle\hat\phi_b|\psi_b^s\rangle$'' for each $\psi_b^s$ in ${\cal
F}_b$. To do this, we first observe that, if we replace $\hat\phi$ in
(\ref{blimq}) by a classical solution, $\phi$, and expand $\phi$ as in
(\ref{phibulkexpansion}), then, by (\ref{phibsym}) and (\ref{eebee}), we
have
\begin{equation}
\label{boundconv}
\phi_b=E_b*\phi_b^s 
\end{equation}
and hence, for all $\phi\in S$ with $\mathfrak{h}(\phi)=\phi_b^s$ and with
boundary-limit $\phi_b$ and for any $\psi\in S$ with
$\mathfrak{h}(\psi)=\psi_b^s$, we have, by (\ref{boundconv}) and the fact that
$\mathfrak{h}$ is a symplectic isomorphism, that
$\langle\phi_b|\psi_b^s\rangle=\langle E_b*\phi_b^s|\psi_b^s\rangle =
-\sigma_b(\phi_b^s, \psi_b^s)= -\sigma(\phi, \psi)$, in view of which
the appropriate definition is clearly 
\begin{equation}
\label{quantb}
\langle\hat\phi_b|\psi_b^s\rangle = -\sigma(\hat\phi, \psi).
\end{equation}
If we now choose \cite{ftnt1} a Poincar\'e chart, ${\cal P}$ and
temporarily adopt the convention of equating any $\psi_b^s\in {\cal
F}_b$ with its restriction to ${\cal P}_b$, (\ref{quantb}) amounts to
saying: \textit{The boundary-limit quantum field $\hat\phi_b$,
``spacetime smeared'' on ${\cal P}_b$ with the test-function $\psi_b^s$,
is equal to minus the ``symplectic smearing'' of the bulk quantum field
$\hat\phi$ with the bulk test-solution $\psi$.}

In view of the fact that $\mathfrak{h}$ is a symplectic isomorphism, the
algebra ${\cal A}_b$ of ``smeared boundary fields'' generated by  the
$\langle\hat\phi_b|\psi_b^s\rangle$ as $\psi_b^s$ ranges over ${\cal
F}_b$ is isomorphic to the bulk field algebra ${\cal A}_B$ generated by
the $\sigma(\hat\phi, \psi)$ as $\psi$ ranges over $S$. (For more
details, see the definition of the ``minimal field algebra'' in
\cite{Kay:2006jn} and note also the options discussed there for
technically different alternatives.)  Moreover, by (\ref{sym_bslick}) 
and (\ref{CCR}), we have 
\begin{equation}
\label{comm}
[\langle\hat\phi_b|{\psi_b^s}_1\rangle,
\langle\hat\phi_b|{\psi_b^s}_2\rangle]=i\langle{\psi_b^s}_1|E_b*{\psi_b^s}_2 
\rangle I
\end{equation} 
and thus (cf. the note after (\ref{eebee})) the subalgebra of ${\cal
A}_b$ generated by test functions in ${\cal F}_b|_{{\cal P}_b}$ may be naturally
identified, when ${\cal P}_b$ is identified with $d$-dimensional
Minkowski space, as the subalgebra of the usual field algebra, ${\cal
A}_d^\Delta$, for the conformal field $\hat\phi_d^\Delta$ obtained by
restricting smearing functions from all of $C^\infty({\cal P}_b)$ to
${\cal F}_b$.

Next we notice that, still for our models (i.e. involving the bulk KG
equation and integer or half-integer $\Delta$) one can, as usual (cf.
\cite{Kay:2006jn}), define a subalgebra ${\cal A}_B({\cal O})$ of our
bulk field algebra for each open region, $\cal O$, of bulk AdS by (cf.
\cite{Kay:2006jn}) taking the algebra generated by the
$\sigma(\hat\phi,\psi)$ where $\psi\in S$ takes the form $E*F$ where $F$
ranges over smooth functions with compact support in $\cal O$. So, in
particular, we obtain an algebra, ${\cal A}_B({\cal W})$, for each bulk
wedge $\cal W$ in AdS (and similarly we obtain an algebra for each bulk
double-cone).  Next we observe that, by (\ref{quantb}), each such bulk
algebra ${\cal A}_B({\cal O})$ is equal to the subalgebra of the
boundary algebra, ${\cal A}_b$, generated by
$\langle\hat\phi_b|\psi_b^s\rangle$ for $\psi_b^s= \mathfrak{h}\psi$,
$\psi=E*F$, $F\in {\cal O}$.  If one makes a choice of Poincar\'e chart,
$\cal P$, then, when $\cal O$ is a wedge, $\cal W\subset {\cal P}$, we
call the latter subalgebra ${\cal A}_b({\cal I})$ where  ${\cal
I}\subset {\cal P}_b$ is the double-cone to which $\cal W$ bijects under
the Rehren bijection \cite{ftnt1}.  In other words, ${\cal A}_b({\cal
I})$ coincides with the element labelled by the region $\cal I$ of the
net of local boundary algebras which gets identified with the element
labelled by the region $\cal W$ of the net of local bulk algebras by the
algebraic holography identification mentioned in our introductory
paragraphs (and similarly for bulk double-cones in ${\cal P}$ and the
boundary regions in ${\cal P}_b$ to which they biject \cite{ftnt2}).  So
in this way our models provide concrete examples of algebraic
holography.  Moreover, in view of the above identification of 
${\cal A}_b$ with ${\cal A}_d^\Delta$, this net of local boundary
algebras may be regarded as a net of local algebras for the conformal
field $\hat\phi_d^\Delta$, but we emphasize \cite{ftnt4} that this
differs from the usual net of local algebras for this theory, not only
because the smearing functions are restricted to elements of ${\cal
F}_b$ but also because these elements are differently ``localized''.

Turning to the connection with boundary-limit holography, if $|0\rangle$
is the Avis et al \cite{Avis:1977yn} ground state for the bulk theory,
i.e. the quasi-free state with symplectically smeared two-point function
$\langle 0|\sigma(\hat\phi,\psi_1)\sigma(\hat\phi, \psi_2)|0\rangle
=\sum_{nl\vec{m}}a_{nl\vec{m}}^{1*}a_{nl\vec{m}}^2$ (where
$a_{nl\vec{m}}^1$ is related to $\psi_1$ as in (\ref{phibulkexpansion})
etc.) then one can show by (\ref{phibsym}) and (\ref{quantb}), that
(again choosing a ${\cal P}$ and re-adopting our convention (see after
(\ref{quantb})) and moreover identifying ${\cal P}_b$ with Minkowski
space) that the ``spacetime-smeared 2-point function'' on the boundary
$\langle 0|(\langle\hat\phi_b|{\psi_b^s}_1\rangle\langle\hat\phi_b|
{\psi_b^s}_2\rangle)|0\rangle$ for a pair of test functions, 
${\psi_b^s}_1, {\psi_b^s}_2 \in {\cal F}_b$, is equal to 
$W_b({\psi_b^s}_1, {\psi_b^s}_2)$ and similarly for all $n$-point
functions.

In view of the fact \cite{Kay:2006jn} that the covariantly smeared bulk
field $\hat\phi(F)$, $F\in C_0^\infty({\rm AdS})$ is equal to the
symplectically smeared field $\sigma(\hat\phi,E*F)$, we conclude from
(\ref{quantb}) that, in our KG models and for $\Delta$ an integer or
half-integer, the bulk smeared Wightman function 
$W(F_1, \dots, F_n)$ is equal to the boundary smeared Wightman
function\hfil\break
$W_b(\mathfrak{h}(-E*F_1), \dots, \mathfrak{h}(-E*F_n))$.  Thus
we see that the test function map $F\mapsto \mathfrak{h}(-E*F)$ induces a
``quantum duality'' between the sets of Wightman functions in bulk and
boundary which are related by Bertola et al's boundary-limit holography.
But in this duality, the test functions with which one smears the
Boundary Wightman functions are restricted to belong to our family 
${\cal F}_b$ (= ${\rm ran}(\mathfrak{h})$). 

Aside from its applications, given in this paper, to providing examples
of algebraic holography and to clarifying its relationship to
boundary-limit holography, we expect that our pre-holography map will be
of use in elucidating other aspects of the AdS/CFT correspondence,
albeit it is only of immediate relevance to the case of bulk theories
which are linear. Furthermore, there are two specific further
conclusions which immediately flow from our results which may be of
relevance to less trivial holography models.  First, that, if one wishes
to construct models on the AdS boundary by requiring them to be related
to the bulk theory by algebraic holography, then this may lead to a more
restricted family of models (in the case of bulk KG, we found only
models with integer or half-integer $\Delta$) than the family one would
obtain by requiring only that the boundary theory be related to the bulk
theory by boundary-limit holography (which, for our bulk KG, have
unrestricted $\Delta$).  The second conclusion concerns the
sometimes-expressed expectation that it is unlikely there could be a
duality between ``ordinary'' QFTs in bulk and boundary because (it is
sometimes said) the boundary having lower dimensions, one would expect
it only to be able to support ``fewer degrees of freedom''.  Surprisingly,
we have found that essentially the opposite to the above expectation
holds true.  Indeed we found that our bulk theory (i.e. AdS$_{d+1}$ KG for
an appropriately tuned mass) is dual to a {\it subtheory} of our
boundary theory -- i.e. to the theory of $\hat\phi_d^\Delta$ after its
test functions have been restricted to the space ${\cal F}_b$.  Our result thus shows us that,
there is, in fact, no simple correlation between dimension and ``degrees
of freedom''.

BSK acknowledges partial support from PPARC.  PL thanks PPARC for a
research studentship.  We thank Karl-Henning Rehren, Bert Schroer and
Atsushi Higuchi for comments on earlier versions of this paper.


\begin{thebibliography}{99}

\bibitem{Maldacena:1997re}
  J.~M.~Maldacena,
  Adv.\ Theor.\ Math.\ Phys.\  {\bf 2}, 231 (1998)
  [Int.\ J.\ Theor.\ Phys.\  {\bf 38}, 1113 (1999)]
  [arXiv:hep-th/9711200].

\bibitem{Gubser:1998bc}
  S.~S.~Gubser, I.~R.~Klebanov and A.~M.~Polyakov,
  Phys.\ Lett.\  B {\bf 428}, 105 (1998)
  [arXiv:hep-th/9802109].

\bibitem{Witten:1998qj}
  E.~Witten,
  Adv.\ Theor.\ Math.\ Phys.\  {\bf 2}, 253 (1998)
  [arXiv:hep-th/9802150].

\bibitem{RehAH}  
K.-H. Rehren
\textit{Ann. Henri Poincar\'e} {\bf 1}, 607 (2000) [arXiv:hep-th/9905179];
Phys.\ Lett.\ B \textbf{493}, 383 (2000)\hfil\break
[arXiv:hep-th/9905179].

\bibitem{Bertola:2000pp}
  M.~Bertola, J.~Bros, U.~Moschella and R.~Schaeffer,
  Nucl.\ Phys.\  B {\bf 587}, 619 (2000)
  [arXiv:hep-th/9908140] .
  
\bibitem{Kay:2006jn}
  B.~S.~Kay,
\textit{Encyclopedia of Mathematical Physics} edited by J.-P. Fran\c
coise, G. Naber and S.T. Tsou (Academic [Elsevier], Amsterdam, New York,
London 2006) Vol. 4, p. 202 [arXiv:gr-qc/0601008].

\bibitem{Haag} R.~Haag, \textit{Local Quantum Physics (2nd ed.)} 
(Springer, Berlin 1996).

\bibitem{ftnt1}  Actually, what we shall call the Rehren bijection here
is slightly different from the definition in \cite{RehAH}. \cite{RehAH}
works on the ${\mathbb Z}_2$ quotient of wrapped AdS and the definition
of the bijection in \cite{RehAH} is adapted to this spacetime.  We
prefer to work on the unwrapped unquotiented AdS (i.e. the covering
space of wrapped unquotiented AdS) because we don't want to have to get
involved with quantum field theory either on spacetimes with closed
timelike curves or on non-orientable spacetimes and also so as not to
preclude from the outset non-integer $\Delta$. With this preference, we
find it necessary to alter the definitions slightly and regard the
Rehren bijection as a bijection between bulk wedges belonging to a
single choice of chart, $\cal P$, for Poincar\'e coordinates and the
double-cones belonging to the boundary, ${\cal P}_b$, of that chart
(which is conformal to a single copy of $d$-dimensional Minkowski space)
and we shall here actually regard algebraic holography as a mapping from
a net of local algebras for such a bulk chart to a net of local algebras
on its boundary (i.e. on Minkowski space).  

\smallskip

Even though the resulting version of algebraic holography refers to a
single choice of Poincar\'e chart, we still wish to privilege global
coordinates.  So in particular we define $\mathfrak{h}$ as a map from
all solutions, $S$, on our (unwrapped, unquotiented) AdS to a set of
functions, ${\cal F}_b$, on our full boundary (albeit
$\sigma_b$ is defined by an integral over a choice of ${\cal P}_b$ -- it
doesn't matter which).  One could, alternatively, work instead
throughout on a single Poincar\'e chart ${\cal P}$ (and use Poincar\'e
coordinates) and define $S$ to be the space of solutions on $\cal P$
(note that this would be the ``same'' space as our $S$ under an obvious
identification) and $\mathfrak{h}$ to be a map from this latter $S$ to a
space of functions on the corresponding ${\cal P}_b$.  However, to do so
would result in considerable loss of global perspective. By defining our
pre-holography map globally, we are able, via (\ref{quantb}) (see the
second full paragraph after (\ref{quantb})) to identify the bulk algebra
for any bounded open region in our AdS with a subalgebra of the algebra
${\cal A}_b$ of its boundary albeit the resulting subalgebras won't
necessarily be identifiable as localized in specific regions and/or (in
the case of bulk wedges and bulk double-cones not contained in $\cal P$) may
be so identifiable but not in a unique way.

\smallskip

We should also point out that our altered definition of ``Rehren
bijection'' misses out a counterpart to the important property
\cite{RehAH} that (when the notion of boundary double-cone is generalized by
adding a conformal boundary to Minkowski space) causally complementary bulk wedges map
to causally complementary boundary double-cones  and, related to this,
our notion of ``algebraic holography'' here is not taken to include a
suitable counterpart to Rehren's property that algebras for
causally complementary bulk wedges get identified with algebras for such
causally complementary boundary double-cones.   Nevertheless it is clear from the
second full paragraph after (\ref{quantb}) that the algebras for causally
complementary bulk wedges do get identified with commuting subalgebras
of ${\cal A}_b$ albeit the identification (cf. the last sentence of the
previous paragraph) of the latter with the algebras for particular
regions of the boundary will be ambiguous (and/but the identification will
be choosable so that the regions are causally complementary).

\bibitem{ftnt2} The Rehren bijection as we define it (see \cite{ftnt1})
extends to a map from the set of all wedges and bulk double-cones in a
choice of Poincar\'e chart ${\cal P}$ on the bulk to the set of
double-cones and certain regions of its boundary, ${\cal P}_b$, the
image of a given bulk double-cone being defined to be the set of points
in the boundary which can be reached by null-geodesics emanating from
it.  We note that such an extension of the Rehren bijection is not made
in \cite{RehAH} but, if a given bulk double-cone (in a given choice of
$\cal P$) is regarded as an intersection of bulk wedges (some of which
will of course not lie in $\cal P$) then the  algebra for the boundary
region to which the given bulk double-cone bijects is, on our
definition, easily seen to coincide, in the class of examples we obtain
here, with the  intersection of the algebras of all those bulk wedges.
By virtue of this together with what is written in \cite{RehAH} one can see
that, in our examples, the resulting
extension of the notion of algebraic holography from the set of bulk
wedges to the set of bulk wedges together with  bulk double-cones is the
appropriate counterpart (i.e. when one unwraps and  unidentifies as
explained in \cite{ftnt1}) to the extension performed in \cite{RehAH}.
 

\bibitem{Avis:1977yn}
  S.~J.~Avis, C.~J.~Isham and D.~Storey,
  Phys.\ Rev.\  D {\bf 18}, 3565 (1978).
  
\bibitem{ftnt3}  Our results also hold in $d=1$ although this case needs
separate treatment.  See P.~Larkin, ``Pre-Holography'' PhD thesis,
University of York (2008).  $\rho$ now goes from $-\pi/2$ to $\pi/2$ and
the boundary consists of two lines.  In place of
(\ref{phibulkexpansion}), (\ref{phibsym}) and (\ref{eebee}), one has
expansions involving Gegenbauer polynomials.  Another difference is
that, in $d=1$, the boundary modes are incomplete (see paragraph after
equation (\ref{sym_bslick})) by only a finite number of missing modes,
whereas for $d\ge 2$ there are a countable number of missing modes. The $d=1$,
$m=0$ ($\Delta=1$) case is particularly simple and instructive.  Here
(see figure 1) it is useful to change coordinates to $(t,x)$ where
$x=\rho+\pi/2$.  Then $S$ consists of functions of form
$\phi(t,x)=f(t+x)-f(t-x)$, where $f$ is periodic with period $2\pi$ and
is fixed uniquely by demanding that it be the derivative of another periodic
function, $\mathfrak{h}$ has the simple closed-form action $\phi\mapsto
f$ and $\sigma_b(f_1,f_2)=2\int_{{\cal P}_b} f_1(t)f_2'(t)dt$ where
${\cal P}_b$ is coordinatized as the interval $(-\pi, \pi)$.  Moreover,
$E_b=2\delta'$, and ${\cal F}_b$ ($={\rm ran}(\mathfrak{h})$) consists
of smooth periodic functions with period $2\pi$ which arise as
derivatives of other such functions.  

\bibitem{Balasubramanian:1999ri}
  V.~Balasubramanian, S.~B.~Giddings and A.~E.~Lawrence,
  JHEP {\bf 9903}, 001 (1999)
  [arXiv:hep-th/9902052].

\bibitem{Bateman} \textit{Higher Transcendental Functions} (Bateman
Manuscript Project), edited by A. Erd\'elyi (McGraw-Hill, New York 1953)
Vols. I-III. 

\bibitem{ftnt4} In our net of local algebras for the conformal field
$\hat\phi_d^\Delta$, our criterion for a smeared field
$\langle\hat\phi_b|\psi_b^s\rangle$ to be localized in a given boundary
double-cone $\cal I$ entails that $\psi_b$ (which, by (\ref{boundconv})
is equal to $E_b*\psi_b^s$) is supported in $\cal I$  rather than the
smearing function $\psi_b^s$ itself.  We have shown that, for certain
values of $d$ and $\Delta$, $\psi_b^s$ will not be supported in $\cal I$.
It is hoped to discuss this further elsewhere. (We conjecture that it will
always be supported in the union of the causal future and the causal
past of $\cal I$.)  Nevertheless, our net always satisfies commutation at
spacelike separation, as can be seen by combining (\ref{comm}),
(\ref{symplecticpi}) and (\ref{sym_bslick}) together with  the fact that
boundary double-cones are spacelike related whenever the bulk wedges to
which they Rehren-biject are spacelike related, and the fact
\cite{Kay:2006jn} that $\sigma(E*F_1, E*F_2)$ ($=\langle
F_1|E*F_2\rangle_{L^2({\cal P}_b}$) vanishes whenever the supports of
$F_1$ and $F_2$ are spacelike related.

\end{thebibliography}
\end{document}